\documentclass{PoS}

\title{Distribution amplitude for the photon-pion transition}

\ShortTitle{TDAs for $\gamma\to\pi$ transition}

\author{\speaker{A.~Courtoy}, S.~Noguera\\%\thanks{A footnote may follow.}\\
        Departamento de F\'{\i}sica Te\'orica and Instituto de F\'{\i}sica Corpuscular,\\
Universidad de Valencia-CSIC,\\ E-46100 Burjassot (Valencia), Spain.\\
        E-mail: \email{aurore.courtoy@uv.es},  \email{santiago.noguera@uv.es}}

\abstract{
The exclusive production of $\pi\pi$ and $\pi\rho$ in hard $\gamma^{\ast}\gamma$ scattering in the 
forward kinematical region where the virtual photon is highly
off-shell are studied through the $\gamma\to\pi^{-} \,$ Transition Distribution Amplitudes. The calculation 
is based on a 
covariant Bethe-Salpeter approach, applied to the Nambu~-~Jona-Lasinio model,  for the determination 
of the pion bound state. In particular  it is shown that the pion pole contribution produces 
a large enhancement of the differential cross section for the pion pair production with respect 
to previous estimates. }

\FullConference{LIGHT CONE 2008 Relativistic Nuclear and Particle Physics\\
		 July 7-11  2008\\
		 Mulhouse, France}

\begin{document}

\begin{figure}[h]
\begin{center}
\includegraphics[height=5cm]{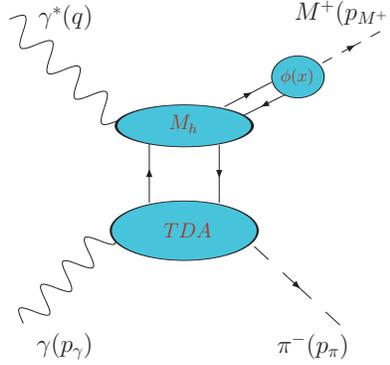}%
\caption{Factorization for the amplitude of the process $\gamma^{\ast}\gamma\rightarrow\pi^- M^+$ at
small momentum transfer.}%
\label{facto}%
\end{center}
\end{figure}

The study of the exclusive meson pair production in $\gamma^{\ast}\gamma$ scattering allows the 
introduction of a new kind of distribution
 amplitudes~\cite{Pire:2004ie}. At small momentum transfer $t$ and in the kinematical regime where the 
 photon is highly virtual, a factorization between the perturbative and 
the nonperturbative regimes is assumed to be  valid. The 
 amplitude for such reactions,  represented in Fig.~\ref{facto}, can be written as
 a convolution of a hard part $M_h$,  with a meson distribution amplitude $\phi_{M}$ and a soft part 
 describing the photon-pion transition. This soft part is called Transition Distribution Amplitude 
 (TDA). 

Cross section estimates for the processes
\begin{equation}
\gamma_{L}^{\ast}\gamma\rightarrow\pi^{+}\pi^{-}\ \ ,\quad\gamma_{L}^{\ast
}\gamma\rightarrow\rho^{+}\pi^{-} \label{processes}%
\end{equation}
have been proposed in Ref.~\cite{Lansberg:2006fv} using for the TDA a
$t$-independent double distributions, in a first approach, and, in a second,
the $t$-dependent results of Ref.~\cite{Tiburzi:2005nj}.
In this paper we evaluate these cross sections in our formalism.
To do so we compare results for the TDAs calculated in different realistic models for the pion
\cite{Broniowski:2007fs, Courtoy:2007vy,Kotko:2008gy}. Since the results obtained are in 
agreement, we choose to use the results of a single model calculation, i.e. the 
NJL model \cite{ Courtoy:2007vy}.
 As shown in Ref.~\cite{pionpole}, the estimate for the $\pi\pi$ production 
 increases by a factor about 60 due to the presence of the pion pole contribution in our analysis.

\section{Transition Distribution Amplitudes}

We introduce the light-front vectors $\bar{p}^{\mu}=P^{+}\left(
1,0,0,1\right)  /\sqrt{2}$ and $n^{\mu}=\left(  1,0,0,-1\right)  /(
\sqrt{2}P^{+})$ where $P^{+}$ is the plus\footnote{We introduce the 
light-cone coordinates $v^{\pm}=\left(  v^{0}\pm v^{3}\right)  /\sqrt{2}$ and
the transverse components $v^{\bot}=\left(  v^{1},v^{2}\right)  $ for
any four-vector $v^{\mu}$. } componente of the vector  $P=\left(  p_{\pi}+p_{\gamma}\right)
/2$.
The momentum transfer is defined as $\Delta=p_{\pi}-p_{\gamma}$, with  $t=\Delta^{2}$  and 
 $P^{2}=m_{\pi}^{2}/2-t/4$. The skewness variable
describes the loss of plus momentum of the incident photon, i.e. $\xi=\left(
p_{\gamma}-p_{\pi}\right)  ^{+}/2P^{+},$ and its value ranges between
$-1<\xi<-t/\left(  2m_{\pi}^{2}-t\right)  $. With these conventions, the vector and axial TDA are
defined by
%\bw%
\begin{eqnarray}
&&\int\frac{dz^{-}}{2\pi}e^{ixP^{+}z^{-}}\left.  \left\langle \pi^{\pm}(p_{\pi
})\right\vert \bar{q}\left(  -\frac{z}{2}\right)  \gamma^{+}\hspace*{0pt}%
\tau^{\pm}\ q\left(  \frac{z}{2}\right)  \left\vert \gamma(p_{\gamma
}\varepsilon)\right\rangle \right\vert _{z^{+}=z^{\bot}=0}\nonumber\\
 &&  =\frac{1}{P^{+}%
}\ i\,e\,\varepsilon_{\nu}\,\epsilon^{+\nu\rho\sigma}\,P_{\rho}\,(p_{\pi
}-p_{\gamma})_{\sigma}\,\frac{V^{\gamma\rightarrow\pi^{\pm}}(x,\xi,t)}%
{\sqrt{2}f_{\pi}}\quad,\nonumber\\
& \label{vectortda}\\
&&\int\frac{dz^{-}}{2\pi}e^{ixP^{+}z^{-}}\left.  \left\langle \pi^{\pm}\left(
p_{\pi}\right)  \right\vert \bar{q}\left(  -\frac{z}{2}\right)  \gamma
^{+}\hspace*{-0.05cm}\gamma_{5}\ \tau^{\pm}\ q\left(  \frac{z}{2}\right)
\left\vert \gamma\left(  p_{\gamma}\varepsilon\right)  \right\rangle
\right\vert _{z^{+}=z^{\bot}=0} \nonumber\\
&&  =\pm\ \frac{1}{P^{+}}\left[  -\ e\,\left(
\vec{\varepsilon}^{\bot}\cdot(\vec{p}_{\pi}^{\bot}-\vec{p}_{\gamma}^{\bot
})\right)  \frac{A^{\gamma\rightarrow\pi^{\pm}}(x,\xi,t)}{\sqrt{2}f_{\pi}%
}\right. % \nonumber\\
%&& 
 \left.  +\ e\,\left(  \varepsilon\cdot(p_{\pi}-p_{\gamma})\right)
\frac{2\sqrt{2}f_{\pi}}{m_{\pi}^{2}-t}~\epsilon(\xi)~\phi_{\pi}\left(
\frac{x+\xi}{2\xi}\right)  \right] 
 .\nonumber\\ &
 \label{axialtda}%
\end{eqnarray}
%\ew
where the pion decay constant is $f_{\pi}=92.4$ MeV, $\epsilon\left(
\xi\right)  $ is equal to $1$ for $\xi>0$ and to $-1$ for $\xi<0$ and
$\phi_{\pi}(x)$ is the pion DA. Here we have modified the definition given in
\cite{Pire:2004ie,Lansberg:2006fv} in order to introduce the pion pole
contribution in the Eq. (\ref{axialtda}) \cite{Tiburzi:2005nj, Courtoy:2007vy}. This pion pole term
describes a point-like pion propagator multiplied by the distribution
amplitude (DA) of an on-shell pion. It contributes to the axial current
through a different momentum structure and must be subtracted in order to
obtain de axial TDA. %We emphasize that this is a model independent definition,
%because we have define the numerator of the pion pole term as the residue at
%the pole $t=m_{\pi}^{2}$. 
With these -model independent- definitions we recover
the sum rules%
\begin{equation}
\int_{-1}^{1}dx\ D\left(  x,\xi,t\right)  =\frac{\sqrt{2}\ f_{\pi}}{m_{\pi}%
}\ F_{D}\left(  t\right)
\ \ \ ,\ \ \ \ \ \ \ \ \ \ \ \ \ \ \ D=V,A\label{sumrules}%
\end{equation}
with the standard definitions for the form factors $F_{V,A}$ appearing in the
$\pi^{\pm}\rightarrow\ell^{\pm}\nu\gamma$ decay \cite{Amsler:2008zz}. Notice that
the on-shell pion DA obeys the normalization condition $\int_{0}^{1}%
dx\,\phi_{\pi}(x)=1$. A model calculation for $A\left(  x,\xi,t\right)  $
implies the evaluation of all diagrams contributing to the matrix element of the axial current 
and to extract
from this result the pion pole contribution calculated in the same model.

\begin{figure}[tb]
\centering
\includegraphics[height=.26\textheight]{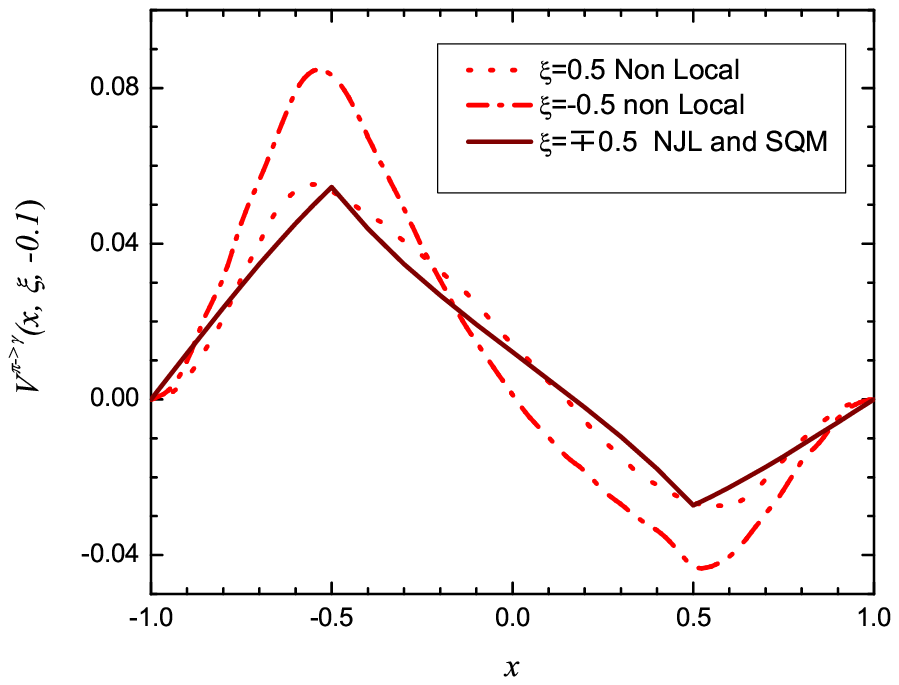}
\includegraphics[height=.28\textheight]{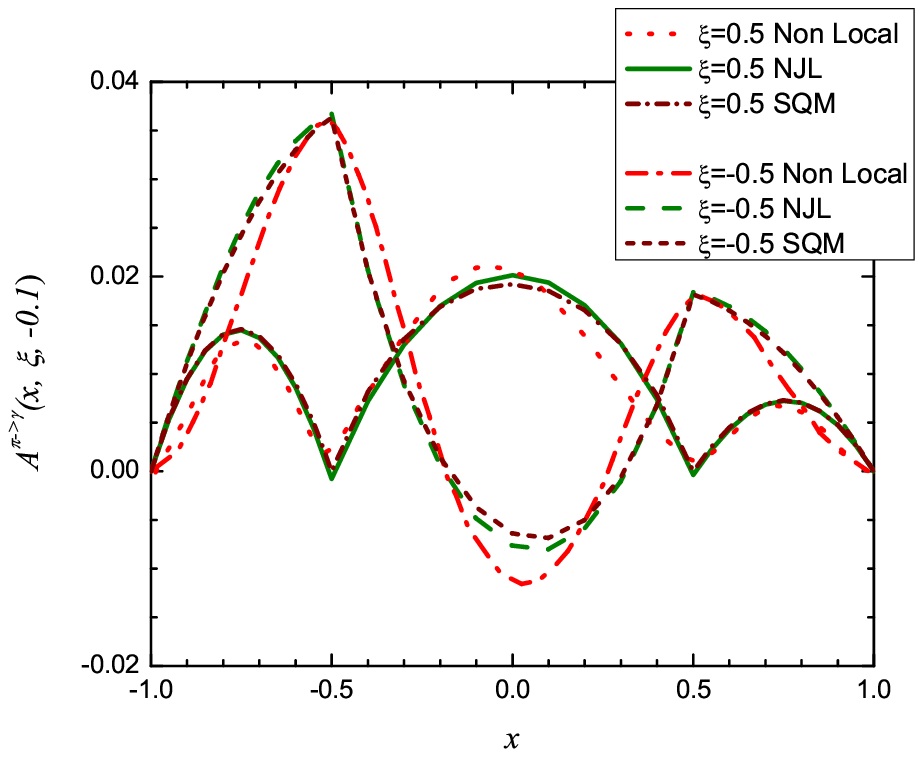}
  \caption{ The $\pi^+$-$\gamma$ TDAs for $t=-0.1$ GeV$^2$ of Refs.~\cite{Broniowski:2007fs, Kotko:2008gy} for $m_{\pi}=0$ MeV  and  Ref.~\cite{Courtoy:2007vy} for $m_{\pi}=140$ MeV.
  On the left,  the vector  TDA  for $\xi=\pm 0.5$ as a single (solid) curve  for the results of both  the NJL model and SQM (these four curves are indistinguishable);
    the result of the non-local $\chi$QM calculation  for $\xi=0.5$ (dotted line) and for $\xi=-0.5$ (dashed-dotted line). On the right, the axial TDA.}
\label{comp}
\end{figure}
Recently the pion-photon TDAs
have been calculated in the
Spectral Quark Model (SQM) \cite{Broniowski:2007fs}, the Nambu - Jona-Lasinio model with
Pauli-Villars regularization procedure (NJL) \cite{Courtoy:2007vy} and a nonlocal chiral
quark model ($\chi$QM) \cite{Kotko:2008gy}. A comparison of the TDAs obtained in  these
three models has been shown in Ref.~\cite{Courtoy:2008ij}. We here recall  the conclusion of the comparison by plotting the different results for the $\pi^+\to\gamma$ transition in Fig.~\ref{comp}.
There is good agreement between the different studies of the pion-photon 
TDAs in spite of the discrepancy coming from the vector TDA for a negative $\xi$ in the non-local $\chi$QM. 
Thus, in the present analysis, we can  concentrate on  the TDAs obtained in a single model, e.g. the NJL
model~\cite{Courtoy:2007vy}. 
The $\gamma$-$\pi$
TDAs defined in Eqs.~(\ref{vectortda}-\ref{axialtda})  
 are connected to the $\pi$-$\gamma$ TDAs through  the $ T$ or
$CPT$ symmetries; we find\footnote{Observe that we have changed
the sign in the definition of $A^{\gamma\rightarrow\pi^{-}}$ with respect to
reference \cite{Courtoy:2008ij}.} \cite{Courtoy:2008ij} 
\begin{eqnarray}
D^{\gamma\rightarrow\pi^{+}}\left(  x,\xi,t\right)  =D^{\pi^{+}\rightarrow\gamma}\left(  x,-\xi,t\right)  \quad \&\quad D^{\gamma\rightarrow\pi^{-}}\left(  x,\xi,t\right)  =D^{\pi^{+}\rightarrow\gamma}\left(  -x,-\xi,t\right) \quad,
\end{eqnarray}
with $D=V,A$.
\section{Exclusive meson production in $\gamma^{\ast}\gamma$ scattering: Cross section estimates}

The $\gamma^{\ast}\gamma\rightarrow M^{+}\pi^{-}$ processes, with $M
=\rho_{L}$ or $\pi$, are subprocesses of the $e\left(  p_{e}\right)
+\gamma\left(  p_{\gamma}\right)  \rightarrow e\left(  p_{e}^{\prime}\right)
+M^{+}\left(  p_{M}\right)  +\pi^{-}\left(  p_{\pi}\right)  $ processes. We
follow all the kinematics given in Section III.~A and Fig. 3 of
Ref.~\cite{Lansberg:2006fv}, but with $n.p=1$. %the exception that our $n^{\mu}$  is
%twice the $n^{\mu}$ vector used in \cite{Lansberg:2006fv} ($n.p=1$ with our
%definitions).
In particular, for massless pions,%
\begin{eqnarray}
&Q^{2}=-q^{2}=-\left(  p_{e}-p_{e}^{\prime}\right)  ^{2}\quad,
\qquad &s_{e\gamma}=\left(  p_{e}+p_{\gamma}\right)  ^{2}\quad,\nonumber\\
&p_{\gamma}=(1+\xi)\bar{p},\quad  & p_{\pi}=(1-\xi)\bar{p}+\frac{\vec{\Delta
}^{\perp2}}{2\left(  1-\xi\right)  }n+\vec{\Delta}^{\perp},\nonumber\\
&&q =-2\xi\bar{p}+\frac{Q^{2}}{4\xi}n\quad,
\end{eqnarray}
where $\Delta_{T}=(0,\vec{\Delta}^{\perp},0)$ and therefore $\Delta_{T}
^{2}=-\vec{\Delta}^{\perp2}$. Notice that $\vec{\Delta}^{\perp2}
=(-t)(1-\xi)/(1+\xi)$, with $t<0$. The longitudinal polarization of the
incoming virtual photon is defined through the conditions $\varepsilon_{L}
^{2}=1$ and that $\varepsilon_{L}.q=0$,%
\[
\varepsilon_{L}=\left(  \frac{2\xi}{Q}\bar{p}+\frac{Q}{4\xi}n\right)  \quad;
\]
while the real photon polarization is defined by $\varepsilon
\cdot p_{\gamma}=0$, which leads to $\varepsilon^{-}=0$ together with the gauge
condition $\varepsilon^{+}=0$.

The differential cross sections are given by\footnote{A
factor of $1/4$ is missing in Eq.~(23) of Ref.~\cite{Lansberg:2006fv}. This
typo does not affect to the numerical results reported there \cite{JPL}.} 
\cite{Lansberg:2006fv}
\begin{eqnarray}
\frac{d\sigma^{e\gamma\to e\rho_L^+\pi^-}}{dQ^2 dt d\xi }&=& \frac{64\pi^2}{9}\,\frac{\alpha_{elm}^3}{s_{e\gamma}\,Q^8}\,(-t)\,\frac{1-\xi}{(1+\xi)^4}
\,\left(2\xi\,s_{e\gamma}\,-\,(1+\xi)Q^2 \right)\,\left \{\Re^2{I}_x^{\rho}+\Im^2 {I}_x^{\rho}\right\}
\label{vfincross};\nonumber\\
\\
\frac{d\sigma^{e\gamma\to e\pi^+\pi^-}}{dQ^2 dt d\xi }&=& \frac{64\pi^2}{9}\,\frac{\alpha_{elm}^3}{s_{e\gamma}\,Q^8}(-t)\frac{1-\xi}{(1+\xi)^4}
\left(2\xi s_{e\gamma}-(1+\xi)Q^2 \right)\left \{\left(\Re {I}_x^{\pi}-\frac{3}{4\pi} 
\frac{Q^{2}\,F_{\pi}\left(Q^{2}\right)}{t-m_{\pi}^2}\right)^2+\Im^2 {I}_x^{\pi}\right\},\nonumber\\
\label{fincross}
\end{eqnarray}
with
\begin{eqnarray}
{I}_x^{\rho}&=& \frac{\alpha_s}{6}
\int_{-1}^1 dx\,\int_0^1 dz\,\left (\frac{f_{\rho}}{\sqrt{2}\,f_{\pi}}\right)\,\phi_{\rho}(z)\,\frac{1}{z (1-z)}\left(\frac{Q_u}{x-\xi+i\epsilon}+\frac{Q_d}{x+\xi-i\epsilon}\right)
 V(x, \xi,t)\quad;
 \label{irho}
\\
{I}_x^{\pi}&=&\frac{\alpha_s}{6}\int_{-1}^1 dx\,\int_0^1 dz\,\phi_{\pi}(z)\,\frac{1}{z (1-z)}\left(\frac{Q_u}{x-\xi+i\epsilon}+\frac{Q_d}{x+\xi-i\epsilon}\right)
 A(x, \xi,t)
 \label{ipi}\quad,
\end{eqnarray}
where $z$ is the light-cone momentum fraction carried by the quark entering
the meson $M^{+},$ $f_{\rho}=0.216$ GeV. The term proportional to $F_{\pi}$ on the r.h.s. of Eq.~(\ref{fincross}) is the pion pole contribution
to the amplitude coming from the second term of Eq.~(\ref{axialtda}).

From Eqs.~(\ref{vfincross}-\ref{fincross}) it can be
observed  that $\xi\geq Q^{2}/\left(2  s_{e\gamma}-Q^{2}\right)$. In other words, there is a (positive) 
lower limit on the value of $\xi$.
 It is indeed a particularly interesting  restriction because  
 the sign of $\xi$ defines  the shape of the axial TDA. 
 %For instance, $V(x,\xi,t) $ has its maximum and minimum values at $x=\pm\xi$ (see Fig.~\ref{comp}). 
 %On the other hand, the $t$-dependence barely controls the magnitude of the distributions. This can be
%easily understood because the -first moments of the- TDAs, that  must satisfy the sum 
%rules Eq.~(\ref{sumrules}),  are 
%expected to decrease at least as $t^{-1}.$ 

We proceed now to the evaluation of the integrals (\ref{irho}-\ref{ipi}). The meson DA, $\phi_{M}\left(
z\right)  ,$ is chosen to be the usual asymptotic normalized meson DA, i.e.
$\phi_{M}(z)=6z(1-z)$, what cancels the $z$-dependence of the hard amplitude.
%The non perturbative part of the process is included in the TDAs. 
Because of the non perturbative information they contain, 
the TDAs have to be evaluated in a  model. We here focus on  the TDAs calculated in the NJL 
model~\cite{Courtoy:2007vy}. This approach is
based on the determination of the pion as a bound state through the
Bethe-Salpeter equation, what guarantees the preservation of all the invariances of
the problem. As a consequence, the
obtained TDAs explicitly verify the sum rules, the
polynomiality condition and have the correct support in $x$. 
The NJL model gives a good
description of the low energy pion physics \cite{Klevansky:1992qe} and it has already been
applied to the study of the pion parton distribution (PD) \cite{Davidson:1994uv}
and the pion generalized parton distribution (GPD) \cite{Theussl:2002xp}. 
Once evolution is taken into account, the calculated PD is in good agreement with the experimental one
\cite{Davidson:1994uv}. The  QCD evolution of the pion GPD calculated in the NJL model
has been studied in \cite{Broniowski:2007si}. More elaborated studies of the pion PD has been done in, e.g., 
the Instanton Liquid Model \cite{Anikin:2000rq}, lattice calculation based
models \cite{Noguera:2005cc} using non local lagrangians \cite{Noguera:2005ej}, which confirms that the result obtained in the NJL model for the PD is a
good approximation. It is therefore of interest to obtain the cross sections
for the processes (\ref{processes}) in such a realistic model.

In order to  numerically estimate the cross sections,  we need to fix the strong coupling constant $\alpha_s$. In Ref.~\cite{Braun:2005be} it is indicated that a large value of $\alpha_{s}$
($\alpha_{s}= 1$) should be used together with the asymptotic DA. We hence use the value $\alpha_{s}= 1$.

\begin{figure}%[tb]
\begin{center}
\includegraphics[height=6.6997cm,width=7.9737cm]{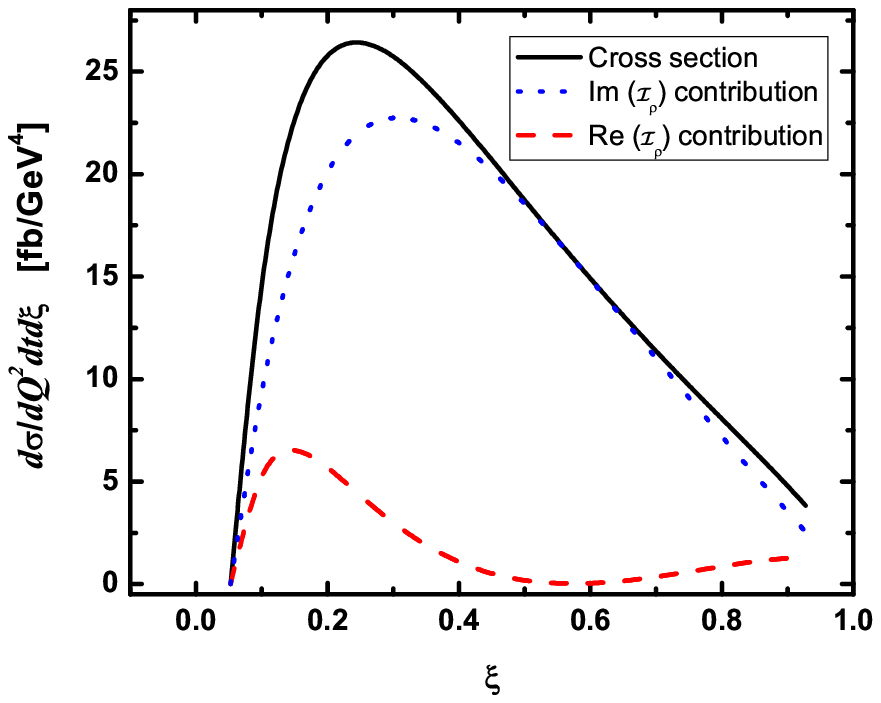}%
\includegraphics[height=6.6997cm,width=7.9737cm]{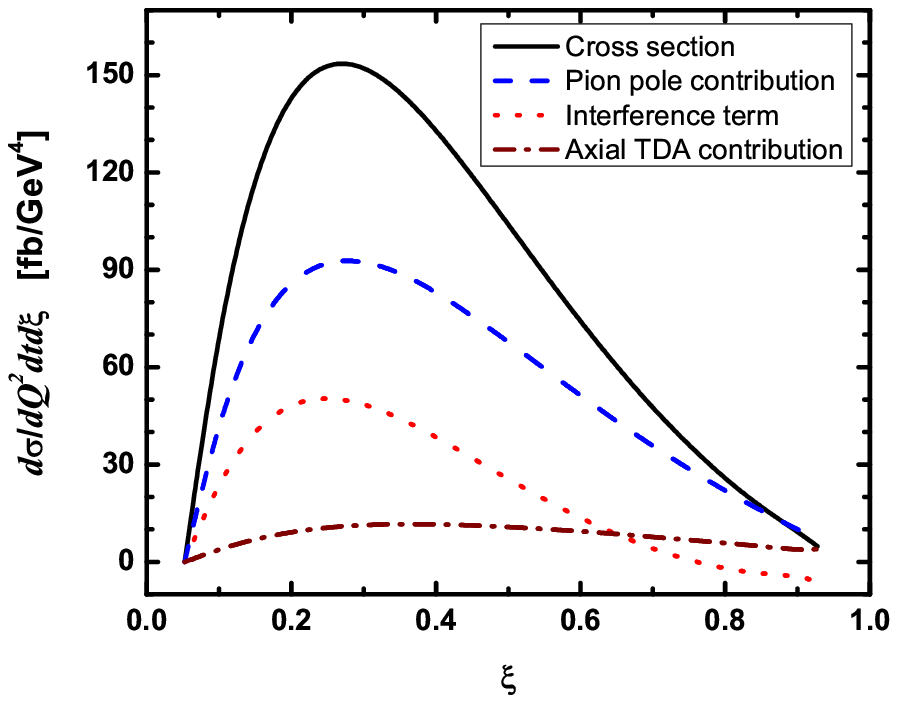}%
\caption{
The $e\gamma\rightarrow e^{\prime}M^{+}\pi^{-}$ differential cross
section with $M=\rho_L$ on the left and $M=\pi$ on the right.  The plots are given as functions of $\xi$ for $Q^{2}=4$ GeV$^{2},$
$s_{e\gamma}=40$ GeV$^{2},$ $t=-0.5$ GeV$^{2}$. %The
%dotted\ (dashed) curve is the contribution to the cross section comming from
%the imaginary (real) part.% of the integral given in Eqs.~(\ref{irho}-\ref{ipi}).
}
\label{pi-rho-4}%
\end{center}
\end{figure}

The result for the cross section for  $\rho$
production  is shown in Fig.~\ref{pi-rho-4} as a function of
$\xi$. %As we observe, 
The cross section is largely dominated by the imaginary
part (dotted line) of the integral of Eq.~(\ref{irho}). Comparing with the previous obtained
results in Ref.~\cite{Lansberg:2006fv}, we observe that our predictions
are higher by a factor 2 or 3.%

The $\pi \pi$ production is described by Eq.~(\ref{fincross}). The pion pole term 
of the axial current leads to a contribution proportional 
to the pion form factor $F_\pi(Q^2)$. If we use the asymptotic 
form of the pion DA, $\phi_{\pi}\left(
z\right)  $ with $z=\left(  x+\xi\right)  /2\xi,$ for the evaluation of this contribution we obtain
the Brodsky-Lepage result for $F_\pi(Q^2)$,
\begin{equation}
\frac{3}{4\pi}Q^{2}\,F_{\pi}\left(Q^{2}\right) =
12\,\alpha_s f_{\pi}^2 \label{pi-FF}
\end{equation}
%with $F_{BL}\left(  Q^{2}\right)  $ the Brodsky-Lepage pion FF
%\begin{equation}
%$Q^{2}\,F_{BL}\left(  Q^{2}\right)  =36\pi\ \alpha_{s}\ f_{\pi}^{2}\, C_F/n_C$. %\quad.
%\end{equation}
The cross section for pion production as a function of $\xi$ is given on the right of
Fig.~\ref{pi-rho-4}. We notice an enhancement of about 2 orders of magnitude  \cite{pionpole} with 
respect to the first estimates given in Ref.~\cite{Lansberg:2006fv}, due to the presence of the pion 
pole in Eq.~(\ref{fincross}). The cross section is indeed dominated by the pion pole
contribution (dashed line) whose behavior is governed by the pion FF.  The latter being experimentally 
determined,  the pion pole contribution is perfectly known. Thus one could extract information about 
the axial TDA from the interference term
(dotted line) whose contribution is more important than the pure axial TDA's contribution (dotted-dashed line).  

We finally remark that, contrary to the contributions coming from the TDAs, the
pion pole strongly depends on the momentum transfer for small values of $t$. Neglecting the pion mass, 
the cross section for this contribution goes like $t^{-1}$. For large values of $t$ the behaviour of
all contributions is as $t^{-1}$.

\acknowledgments
This work has been supported  by the Sixth Framework Program of the
European Commision under the Contract No. 506078 (I3 Hadron Physics); 
by the MEC (Spain) under the Contract FPA 2007-65748-C02-01 and 
the grant AP2005-5331 and by EU FEDER. %Feynman diagrams drawn using JaxoDraw \cite{Binosi:2003yf}.

\end{document}